# Resonant properties of composite structures consisting of several resonant diffraction gratings

LEONID L. DOSKOLOVICH,[1,2,*] EVGENI A. BEZUS,[1,2] DMITRY A. BYKOV,[1,2] NIKITA V. GOLOVASTIKOV,[1,2] AND VICTOR A. SOIFER[1,2]

[1]Image Processing Systems Institute — Branch of the Federal Scientific Research Centre "Crystallography and Photonics" of Russian Academy of Sciences, 151 Molodogvardeyskaya st., Samara 443001, Russia
[2]Samara National Research University, 34 Moskovskoe shosse, Samara 443086, Russia
*leonid@smr.ru

**Abstract:** We theoretically and numerically investigate resonant optical properties of composite structures consisting of several subwavelength resonant diffraction gratings separated by homogeneous layers. Using the scattering matrix formalism, we demonstrate that the composite structure comprising $N$ gratings has a multiple transmittance zero of the order $N$. We show that at the distance between the gratings satisfying the Fabry–Pérot resonance condition, an $(N-1)$-degenerate bound state in the continuum (BIC) is formed. The results of rigorous numerical simulations fully confirm the theoretically predicted formation of multiple zeros and BICs in the composite structures. Near the BICs, an effect very similar to the electromagnetically induced transparency is observed. We show that by proper choice of the thicknesses of the layers separating the gratings, nearly rectangular reflectance or transmittance peaks with steep slopes and virtually no sidelobes can be obtained. In particular, one of the presented examples demonstrates the possibility of obtaining an approximately rectangular transmittance peak having a significantly subnanometer width. The presented results may find application in the design of optical filters, sensors and devices for optical differentiation and transformation of optical signals.

## 1. Introduction

The effect of optical resonance is utilized in a wide class of photonic devices possessing unique optical properties [1]. Diffraction gratings (DGs) constitute one of the most widespread classes of resonant photonic structures. Despite a long history (resonances in DGs were first observed by R. Wood in 1902 [2]), resonant DGs remain the subject of intensive research due to numerous extraordinary optical effects arising in resonant conditions [3, 4].

In recent years, significant research attention was dedicated to the investigation of the so-called bound states in the continuum (BICs) supported by various photonic structures and, in particular, by diffraction gratings [5–10]. Under BICs, a class of the eigenmodes of the structure is understood, which, although coexisting with a continuous spectrum of radiating waves, remain perfectly confined and, therefore, have an infinite quality factor and a real frequency. The leakage of the mode energy to the open scattering channels can be eliminated, for example, by properly choosing the parameters of the structure so that the amplitudes of the outgoing waves vanish due to destructive interference. A small detuning from the BIC condition enables obtaining resonances with extremely high quality factors. This leads to many potential applications of photonic structures supporting BICs, including narrowband filters and sensors.

In the present work, we investigate composite dielectric diffractive structures consisting of several identical resonant DGs with subwavelength period separated by homogeneous dielectric layers. A distinctive feature of resonant subwavelength dielectric gratings is the presence of zeros in the transmittance spectrum [11, 12]. At the corresponding frequencies and angles of incidence, the radiation incident on the grating is totally reflected. Existence of the transmittance zeros makes it possible to use resonant DGs as optical filters [4, 13], sensors [4, 14, 15], and even devices for analog optical differentiation of optical signals

[16, 12]. In this work, using the scattering matrix formalism, we show that a composite structure comprising $N$ identical DGs having a transmittance zero enables obtaining a multiple transmittance zero of the order $N$. It is demonstrated that at the distances between the DGs satisfying the Fabry–Pérot resonance condition, $N-1$ bound states in the continuum, which are degenerate, are formed in the composite structure. In the vicinity of the BICs, an effect very similar to the electromagnetically induced transparency is observed. The presence of several resonances (and BICs) in the composite structure provides additional opportunities for controlling the shape of the spectra by choosing the thicknesses of the dielectric layers separating the DGs. In particular, one of the presented examples demonstrates the possibility of obtaining a resonant transmittance peak having a prescribed shape and a substantially subnanometer width.

## 2. Scattering matrix formalism

The optical properties of a diffraction grating can be described by a scattering matrix. The scattering matrix **S** relates the complex amplitudes of the plane waves incident on the grating with the amplitudes of the transmitted and reflected diffraction orders [11, 17–19]:

$$\begin{bmatrix} \mathbf{T} \\ \mathbf{R} \end{bmatrix} = \mathbf{S} \begin{bmatrix} \mathbf{I}_u \\ \mathbf{I}_d \end{bmatrix}, \quad (1)$$

where **R** and **T** are the vectors of complex amplitudes of the reflected and transmitted diffraction orders, respectively, and $\mathbf{I}_u$ and $\mathbf{I}_d$ are the vectors of complex amplitudes of the plane waves impinging on the DG from the superstrate and the substrate regions, respectively. In what follows, we assume that the elements of the scattering matrix of Eq. (2) are functions of the angular frequency $\omega$ of the incident light. Optical properties of a subwavelength DG, in which only the 0[th] reflected and transmitted diffraction orders propagate (are non-evanescent), can be described by a $2\times 2$ scattering matrix

$$\mathbf{S}_1(\omega) = \begin{pmatrix} t_1(\omega) & r_{d,1}(\omega) \\ r_{u,1}(\omega) & t_1(\omega) \end{pmatrix}, \quad (2)$$

where $t_1(\omega)$ is the complex transmission coefficient of the DG (complex amplitude of the 0[th] transmitted diffraction order) for a unit-amplitude wave incident from the superstrate or the substrate region, and $r_{u,1}(\omega)$ and $r_{d,1}(\omega)$ are the complex reflection coefficients (complex amplitudes of the 0[th] reflected diffraction orders) for unit-amplitude waves impinging on the DG from the superstrate and the substrate regions, respectively. It is worth noting that the scattering matrix of Eq. (2) does not describe the near-field effects, which are associated with the evanescent diffraction orders of the DG. In addition, let us mention that the matrix of Eq. (2) also allows one to describe the optical properties of multilayer diffractive structures containing homogeneous layers and subwavelength diffraction gratings.

Let us consider a composite structure (composite DG) consisting of two subwavelength DGs described by scattering matrices of Eq. (2) and separated by a homogeneous dielectric layer with the thickness $l$ and the refractive index $n$ (Fig. 1). In this case, the scattering matrix of the composite DG can be expressed through the matrix $\mathbf{S}_1(\omega)$ in the form [17, 19, 20]

$$\mathbf{S}_2(\omega) = \mathbf{S}_1(\omega) * \mathbf{L}(\omega) * \mathbf{S}_1(\omega), \quad (3)$$

where the symbol $*$ denotes the Redheffer star product [17] defined as

$$\begin{pmatrix} a_{1,1} & a_{1,2} \\ a_{2,1} & a_{2,2} \end{pmatrix} * \begin{pmatrix} b_{1,1} & b_{1,2} \\ b_{2,1} & b_{2,2} \end{pmatrix} = $$
$$= \frac{1}{1-a_{1,2}b_{2,1}} \begin{pmatrix} b_{1,1}a_{1,1} & b_{1,2} - a_{1,2}(b_{1,2}b_{2,1} - b_{1,1}b_{2,2}) \\ a_{2,1} - b_{2,1}(a_{1,2}a_{2,1} - a_{1,1}a_{2,2}) & a_{2,2}b_{2,2} \end{pmatrix}, \quad (4)$$

and $\mathbf{L}(\omega)$ is the scattering matrix of the homogeneous dielectric layer. Upon the propagation through this layer, the plane waves corresponding to the $0^{\text{th}}$ diffraction orders acquire only the phase shift

$$\psi(\omega) = l\sqrt{(n\omega/c)^2 - k_x^2(\omega)}, \quad (5)$$

where $k_x(\omega) = (\omega/c) n_{\text{env}} \sin\theta$ is the tangential component of the wavevectors of the waves impinging on the composite structure from the superstrate and the substrate regions, $c$ is the speed of light in vacuum, $\theta$ is the angle of incidence, and $n_{\text{env}}$ is the refractive index of the surrounding medium. Thus, the matrix $\mathbf{L}(\omega)$ reads as

$$\mathbf{L}(\omega) = \exp\{i\psi(\omega)\} \mathbf{E}, \quad (6)$$

where $\mathbf{E}$ is the $2\times 2$ identity matrix. By substituting Eqs. (2) and (6) into Eq. (3), we obtain the scattering matrix of the composite DG in the form

$$\mathbf{S}_2(\omega) = \begin{pmatrix} t_2(\omega) & r_{d,2}(\omega) \\ r_{u,2}(\omega) & t_2(\omega) \end{pmatrix} = \frac{1}{1 - \exp\{2i\psi\} r_{u,1} r_{d,1}} \cdot \\ \cdot \begin{pmatrix} \exp\{2i\psi\} t_1^2 & r_{d,1}\left[1 - \exp\{2i\psi\}(r_{u,1} r_{d,1} - t_1^2)\right] \\ r_{u,1}\left[1 - \exp\{2i\psi\}(r_{u,1} r_{d,1} - t_1^2)\right] & \exp\{2i\psi\} t_1^2 \end{pmatrix}. \quad (7)$$

The obtained Eqs. (3)–(7) are convenient for the analysis of the optical properties of composite structures carried out in the rest of the present work.

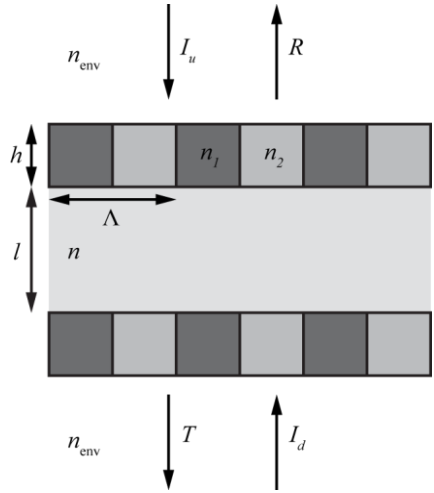

Fig. 1. Geometry of a composite structure containing two identical diffraction gratings with period $\Lambda$, height $h$, and refractive indices $n_1$ and $n_2$ separated by a dielectric layer with the thickness $l$ and refractive index $n$. The structure is located in a symmetric dielectric environment with refractive index $n_{\text{env}}$. The arrows denote the incident, reflected and transmitted waves.

## 3. High-order zeros of the transmission coefficient of composite structures

Let us prove that if the initial DG has a transmission coefficient zero at a certain frequency $\omega = \omega_0$ (i.e. $t_1(\omega_0) = 0$, $|r_{(u,d),1}(\omega_0)| = 1$), then the composite structure described by Eq. (3) will have a transmission coefficient zero of the second order. We start by expanding the elements of the scattering matrix of Eq. (2) into Taylor series in the vicinity of the frequency $\omega = \omega_0$ up to linear terms:

$$t_1(\omega) = \tau(\omega - \omega_0) + \mathrm{O}(\omega - \omega_0)^2,$$
$$r_{u,1}(\omega) = \exp\{i\varphi_u\} + \rho_u(\omega - \omega_0) + \mathrm{O}(\omega - \omega_0)^2, \quad (8)$$
$$r_{d,1}(\omega) = \exp\{i\varphi_u\} + \rho_d(\omega - \omega_0) + \mathrm{O}(\omega - \omega_0)^2.$$

Let us also expand the function $\exp\{i\psi(\omega)\}$ in Eq. (6) up to the linear term:

$$\exp\{i\psi(\omega)\} = \exp\{i\psi_0\} + \gamma(\omega - \omega_0) + \mathrm{O}(\omega - \omega_0)^2, \quad (9)$$

where $\psi_0 = \psi(\omega_0)$. Substituting the expansions (8) and (9) into Eq. (7), after some simple transformations we obtain the elements of the "composite" scattering matrix $\mathbf{S}_2(\omega)$ in the form

$$t_2(\omega) = \frac{\exp\{i\psi\}\tau^2}{1 - \exp\{2i(\psi_0 + \varphi)\}}(\omega - \omega_0)^2 + \mathrm{O}(\omega - \omega_0)^3,$$
$$r_{(u,d),2}(\omega) = \exp\{i\varphi_{(u,d)}\} + \rho_{(u,d)}(\omega - \omega_0) + \mathrm{O}(\omega - \omega_0)^2, \quad (10)$$

where $\varphi = (\varphi_u + \varphi_d)/2$. Thus, we obtained that the transmission coefficient of the composite structure has a zero of the second order, whereas the expansions of the reflection coefficients $r_{(u,d),2}$ still contain both the zero- and first-order terms. Let us note that the representations given by Eq. (10) cannot be used if the condition

$$\psi_0 + \varphi = \pi m, \; m \in \mathbb{N} \quad (11)$$

holds. The condition (11) describes the formation of the Fabry–Pérot resonance between the two DGs. In this case, the denominator of the fraction in the expression for the coefficient $t_2(\omega)$ in Eq. (10) vanishes. By assuming in Eq. (9) that $\exp\{i\psi_0\} = \exp\{i(\pi m - \varphi)\}$ and carrying out a derivation similar to the derivation of Eq. (10), one can easily show that under the condition of Eq. (11), the transmission coefficient of the composite structure has only a first-order zero:

$$t_2(\omega) = -\frac{\exp\{-i\varphi\}\tau^2}{2\exp\{3i\varphi\} + (-1)^m \left(\exp\{i\varphi_d\}\rho_u + \exp\{i\varphi_u\}\rho_d\right)}(\omega - \omega_0) + \mathrm{O}(\omega - \omega_0)^2. \quad (12)$$

In what follows, let us show that a composite structure consisting of $N$ DGs described by the scattering matrix of Eq. (2) has a transmission zero of the order $N$. This can be easily proven by induction. Indeed, at $N = 2$, this statement has already been proven. Let us now show that if a composite structure described by the scattering matrix $\mathbf{S}_N(\omega)$ has a zero of the order $N$, then the composite structure described by the matrix

$$\mathbf{S}_{N+1}(\omega) = \mathbf{S}_N(\omega) * \mathbf{L}(\omega) * \mathbf{S}_1(\omega) \quad (13)$$

will have a zero of the order $N + 1$. Let the elements of the "composite" scattering matrix $\mathbf{S}_N(\omega)$ have the form

$$t_N(\omega) = \frac{\exp\{i(N-1)\psi\}\tau^N}{\left[1 - \exp\{2i(\psi_0 + \varphi)\}\right]^{N-1}}(\omega - \omega_0)^N + \mathrm{O}(\omega - \omega_0)^{N+1},$$
$$r_{(u,d),N}(\omega) = \exp\{i\varphi_{(u,d)}\} + \rho_{(u,d)}(\omega - \omega_0) + \mathrm{O}(\omega - \omega_0)^2. \quad (14)$$

Let us note that at $N = 2$, Eq. (14) turns into Eq. (10). By substituting Eqs. (8), (9) and (14) into Eq. (13), we obtain the elements of the scattering matrix $\mathbf{S}_{N+1}(\omega)$:

$$t_{N+1}(\omega) = \frac{\exp\{iN\psi\}\tau^{N+1}}{\left[1-\exp\{2i(\psi_0+\varphi)\}\right]^N}(\omega-\omega_0)^{N+1} + O(\omega-\omega_0)^{N+2},$$

$$r_{(u,d),N+1}(\omega) = \exp\{i\varphi_{(u,d)}\} + \rho_{(u,d)}(\omega-\omega_0) + O(\omega-\omega_0)^2.$$  (15)

The obtained expressions (15) show that the composite structure described by the scattering matrix $\mathbf{S}_{N+1}(\omega)$ of Eq. (13) has a transmission zero of the order $N+1$. It is worth noting that the first two terms in the Taylor series expansions of the coefficients $r_{(u,d),N+1}(\omega)$ do not depend on $N$, and, in particular, are equal to the corresponding terms in the expansions for $r_{(u,d),1}(\omega)$ in Eq. (8). As in the case of $N=2$, the expression for $t_{N+1}(\omega)$ in Eq. (15) cannot be used if the Fabry–Pérot condition of Eq. (11) holds. By induction, one can show that in this case the transmission coefficient $t_{N+1}(\omega)$ has only a first-order zero.

## 4. Bound states in the continuum in composite structures

As a rule, the zeros in the transmittance and reflectance spectra of a diffraction grating are associated with the quasiguided eigenmodes supported by the structure. The frequencies of the eigenmodes correspond to the poles of the scattering matrix (the poles of the reflection and transmission coefficients) considered as a function of the complex frequency $\omega \in \mathbb{C}$ [18, 21]. In order to simplify the further analysis, let us assume that the transmittance zero of the initial DG is associated with a resonance having a Lorentzian line shape in reflection. This line shape is often observed in subwavelength dielectric DGs [13]. In the case of a Lorentzian resonance, the elements of the scattering matrix of Eq. (2) can be approximated by the following expressions:

$$r_{u,1}(\omega) = \exp\{i\varphi_u\}\frac{i\,\mathrm{Im}\,\omega_p}{\omega-\omega_p},\ r_{d,1}(\omega) = \exp\{i\varphi_d\}\frac{i\,\mathrm{Im}\,\omega_p}{\omega-\omega_p},$$

$$t_1(\omega) = \exp\{i\varphi\}\frac{\omega-\mathrm{Re}\,\omega_p}{\omega-\omega_p},$$  (16)

where $\omega_p$ is the complex frequency of the eigenmode supported by the DG, and $\varphi = (\varphi_u + \varphi_d)/2$. The form of the reflection and transmission coefficients used in the resonant representations (16) ensures the unitarity of the scattering matrix $\mathbf{S}_1(\omega)$. According to Eq. (16), the transmission coefficient $t(\omega)$ vanishes at $\omega_0 = \mathrm{Re}\,\omega_p$. The width of the resonance is determined by the imaginary part of the pole $\omega_p$ and can be characterized by the quality factor $Q = \mathrm{Re}\,\omega_p/2|\mathrm{Im}\,\omega_p|$. From Eq. (16), it is easy to obtain that the width of the resonant reflectance peak $R(\omega) = |r_{(u,d),1}(\omega)|^2$ (resonant transmittance dip $T(\omega) = |t_1(\omega)|^2$) amounts to $\Delta = 2|\mathrm{Im}\,\omega_p|$ at the 0.5 level.

Next, we consider composite structures comprising DGs with Lorentzian spectra described by Eq. (16). We assume that the phase shift $\psi_0 = \psi(\omega_0)$ acquired upon propagation through the layers separating the gratings is of the order of $2\pi$ at the frequency $\omega = \omega_0$. In this case, the thickness of the dielectric layers is $l \sim 2\pi c/(\omega_0 n) = \lambda_0/n$, where $\lambda_0$ is the free-space wavelength corresponding to the frequency $\omega_0$. It is easy to obtain that under the condition $\omega_0 = \mathrm{Re}\,\omega_p \gg \mathrm{Im}\,\omega_p$, which holds even for resonances with a low quality factor $Q \sim 100$, the function $\exp\{i\psi(\omega)\}$ in Eqs. (5)–(7) can be approximated by a constant $\exp\{i\psi_0\}$. Therefore, in what follows, we consider the scattering matrix of the layer separating the gratings to be frequency-independent:

$$\mathbf{L} = \exp\{i\psi_0\}\mathbf{E} = \exp\left\{i\frac{\omega_0}{c}l\sqrt{n^2 - (n_{\text{env}}\sin\theta)^2}\right\}\mathbf{E}. \tag{17}$$

From Eqs. (7), (15), (16) and (17), it follows that in the general case, a composite structure containing $N$ DGs has a transmission zero of the order $N$ and $N$ poles. At the same time, under the condition of Eq. (11), only one zero and one pole remain. Let us demonstrate this effect for the composite structures consisting of two and three resonant DGs. Substituting the resonant representations (16) into Eq. (7), after some simple transformations we obtain the transmission coefficient of a composite structure consisting of two DGs in the form

$$t_2(\omega) = \exp\{i(\psi_0 + 2\varphi)\}\frac{(\omega - \text{Re}\,\omega_p)^2}{(\omega - \omega_{p,1})(\omega - \omega_{p,2})}, \tag{18}$$

where

$$\begin{aligned}\omega_{p,1} &= \text{Re}\,\omega_p + i\left[1 - \exp\{i(\psi_0 + \varphi)\}\right]\text{Im}\,\omega_p, \\ \omega_{p,2} &= \text{Re}\,\omega_p + i\left[1 - \exp\{i(\psi_0 + \varphi)\}\right]\text{Im}\,\omega_p.\end{aligned} \tag{19}$$

In accordance with the proof of the multiplicity of the transmission zero given above, the considered composite grating has a second-order transmission zero at $\omega_0 = \text{Re}\,\omega_p$ and two poles defined by Eq. (19). Under the condition of Eq. (11), one of the poles becomes real and coincides with the real-valued transmission zero. For the sake of argument, let $\text{Im}\,\omega_{p,1} = 0$. In this case, the integer $m$ in Eq. (11) is even, and $\omega_{p,2} = \text{Re}\,\omega_p + 2i\,\text{Im}\,\omega_p$. The eigenmode of the composite structure having the frequency $\omega_{p,1} = \text{Re}\,\omega_p$ is a bound state in the continuum (BIC) [5–10]. BIC is a mode that coexists with a continuum of radiating waves but has an infinite quality factor (a real frequency). The considered structure consisting of two DGs has only two scattering channels corresponding to the 0[th] reflected and transmitted diffraction orders. The leakage to these channels is canceled if the condition of Eq. (11) holds, which describes the Fabry–Pérot resonance formed between the diffraction gratings constituting the composite structure.

A similar behavior is observed in the composite structure consisting of three DGs. Calculating the scattering matrix $\mathbf{S}_3(\omega) = \mathbf{S}_2(\omega) * \mathbf{L} * \mathbf{S}_1(\omega)$, we obtain the transmission coefficient of this composite structure in the form

$$t_3(\omega) = \exp\{i(2\psi_0 + 3\varphi)\}\frac{(\omega - \text{Re}\,\omega_p)^3}{(\omega - \omega_{p,1})(\omega - \omega_{p,2})(\omega - \omega_{p,3})}, \tag{20}$$

where

$$\begin{aligned}\omega_{p,1} &= \text{Re}\,\omega_p + i(1-\sigma)\text{Im}\,\omega_p, \\ \omega_{p,2} &= \omega_p + i\frac{\sigma - \sqrt{\sigma(8+\sigma)}}{2}\text{Im}\,\omega_p, \\ \omega_{p,3} &= \omega_p + i\frac{\sigma + \sqrt{\sigma(8+\sigma)}}{2}\text{Im}\,\omega_p,\end{aligned} \tag{21}$$

where $\sigma = \exp\{2i(\psi_0 + \varphi)\}$. According to Eq. (20), the transmission coefficient of the composite structure comprising three DGs has a third-order zero and three poles. Under the condition of Eq. (11), the poles defined by Eq. (21) take the form

$$\omega_{p,1} = \omega_{p,2} = \text{Re}\,\omega_p, \quad \omega_{p,3} = \text{Re}\,\omega_p + 3i\,\text{Im}\,\omega_p. \tag{22}$$

Thus, if the condition of Eq. (11) holds, the poles $\omega_{p,1}$ and $\omega_{p,2}$ become identical and real and coincide with the real-valued transmission zeros. This means that in the composite structure with three DGs, two bound states in the continuum exist. Since the BIC frequencies coincide, their arbitrary linear combination is also a bound state; therefore, the BICs supported by the structure described by Eqs. (11), (20)–(22) turn out to be double-degenerate.

Finally, let us discuss the behavior of the composite structure comprising $N$ DGs. As it was shown in Section 2, the composite structure consisting of $N$ DGs has a transmittance zero of the order $N$ and $N$ poles. Therefore, the transmission coefficient of such a composite structure can be written as

$$t_N(\omega) = \exp\{i\gamma\} \frac{(\omega - \operatorname{Re}\omega_p)^N}{\prod_{m=1}^{N}(\omega - \omega_{p,m})}, \quad (23)$$

where $\gamma$ is a certain constant, and $\omega_{p,m}$, $m = 1,...,N$ are the poles of the transmission coefficient. Under the Fabry–Pérot resonance condition of Eq. (11), only a first-order transmission zero and a single pole remain in the composite structure. This means that $N-1$ poles of the composite DG become real and coincide with the real-valued transmittance zero $\omega_0 = \operatorname{Re}\omega_p$. Thus, if the condition of Eq. (11) holds, $(N-1)$-degenerate BICs are formed in the composite structure.

Let us note that the formation of a Fabry–Pérot BIC in the composite structure consisting of two resonant structures is known. In particular, Fabry–Pérot BICs formed in a structure consisting of two DGs were considered in Refs. [6–8]. At the same time, to the best of our knowledge, the effect of the formation of multiply-degenerate BICs in the composite structure consisting of $N$ DGs has not been considered yet. Let us note that the Lorentzian shape of the resonance supported by the single DG is not necessary for the formation of a set of BICs. The representations given by Eq. (16) were chosen solely for the convenience of the analysis of the structures composed of two and three DGs. In addition, for Lorentzian resonances, one can easily prove by induction that if the condition of Eq. (11) holds, the transmission coefficient of the composite structure takes the form

$$t_N(\omega) = (-1)^{N-1} \exp\{i\varphi\} \frac{\omega - \operatorname{Re}\omega_p}{\omega - \operatorname{Re}\omega_p + iN\operatorname{Im}\omega_p}, \quad (24)$$

i.e. the only remaining pole $\omega_{p,N} = \operatorname{Re}\omega_p + iN\operatorname{Im}\omega_p$ has an imaginary part $N$ times greater than that of the pole $\omega_p$ of the transmission coefficient of the initial DG.

## 5. Numerical simulation results

Let us numerically investigate the resonant properties of the considered composite structures. For this, let us first discuss the resonant behavior of the diffraction grating used as a building block of the composite structures.

### 5.1. Resonant diffraction grating

As the initial DG, let us use a single-layer (binary) resonant grating with the parameters taken from Ref. [13]. The reflectance and transmittance spectra of the grating are shown in Fig. 2, and the parameters of the DG are given in the figure caption. The inset to Fig. 2 shows the geometry of the grating. The spectra were calculated using an in-house implementation of the Fourier modal method (also known as rigorous coupled-wave analysis) [17] for the case of normal incidence of a TE-polarized plane wave. Due to the existence of a horizontal symmetry plane of the DG, the coefficients $r_{u,1}(\omega)$ and $r_{d,1}(\omega)$ are equal: $r_{u,1}(\omega) = r_{d,1}(\omega) = r_1(\omega)$. From Fig. 2, it is evident that the spectra have an approximately Lorentzian shape. Rigorous calculation shows that the scattering matrix of the considered DG has a pole with the complex frequency $\omega_p = 3.5863 \cdot 10^{15} - 6.0108 \cdot 10^{12} i\, \text{s}^{-1}$ [21]. The transmission coefficient vanishes at $\omega = \omega_0 = \operatorname{Re}\omega_p$ (at the wavelength $\lambda_0 = 525.2$ nm). At this frequency, $r(\omega_0) = \exp\{-0.0284\mathrm{i}\}$, i.e. $\varphi_{u,1} = \varphi_{d,1} = -0.0284$.

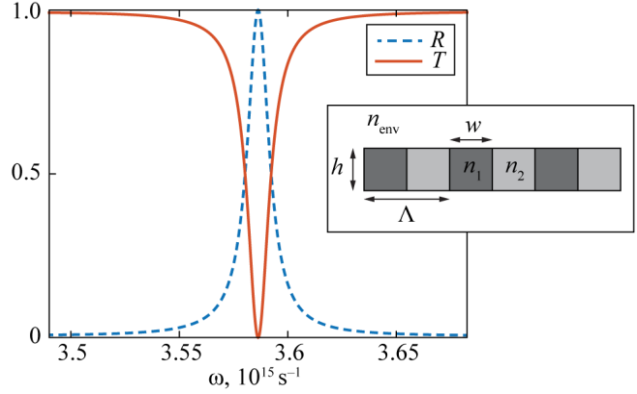

Fig. 2. Spectra $R(\omega) = |r(\omega)|^2$, $T(\omega) = |t(\omega)|^2$ of the resonant diffraction grating with the following parameters: period $\Lambda = 300$ nm, grating height $h = 130$ nm, refractive indices of the grating materials $n_1 = 2.1$ and $n_2 = 1.9$, width of the region with the refractive index $n_1$ $w = \Lambda/2$, refractive index of the surrounding medium $n_{env} = 1.52$. The inset shows the geometry of the grating.

## 5.2. Bound states in the continuum

Figures 3(a) and 4(a) show the reflectance $R_N(\omega, l) = |r_N(\omega, l)|^2$ and transmittance $T_N(\omega, l) = |t_N(\omega, l)|^2$ of the composite structures at $N = 3$ and $N = 4$ vs. the angular frequency of the incident light $\omega$ and the distance $l$ between the DGs. Horizontal dashed lines depict the distances, at which the condition of Eq. (11) is fulfilled, i.e. the Fabry–Pérot resonances are formed. Vertical dashed lines show the frequency $\omega_0 = \operatorname{Re}\omega_p$. In the vicinity of intersections of these lines, resonant features manifested in the transmittance maxima and reflectance minima are clearly visible. When approaching the intersections, the width of the resonances decreases (the quality factor increases). Note that the sharp transmittance peaks arise against the background of a relatively smooth dip, the latter being caused by the existence of a multiple zero of the transmission coefficient. The resulting transmittance spectrum shape is very similar to the spectral shape associated with the electromagnetically induced transparency effect. At the dashed lines, the resonances vanish, which confirms the formation of BICs. The vanishing of the resonant features is clearly demonstrated in the magnified fragments of the transmittance dependence $T_N(\omega, l)$ shown in Figs. 3(c) and 4(c). Let us note that the number of the vanishing resonances in the vicinity of each intersection coincides with the number of the formed BICs (two BICs at $N = 3$ and three BICs at $N = 4$). The vanishing of the resonances is also confirmed by Figs. 3(b) и 4(b), which show the transmittance spectra of the composite structures under the Fabry–Pérot resonance condition depicted with the upper horizontal dashed lines in Figs. 3(a) and 4(a) ($l = 1038$ nm). In Figs. 3(b) and 4(b), only one resonance can be observed, which, according to Eq. (24), has a significantly lower Q-factor.

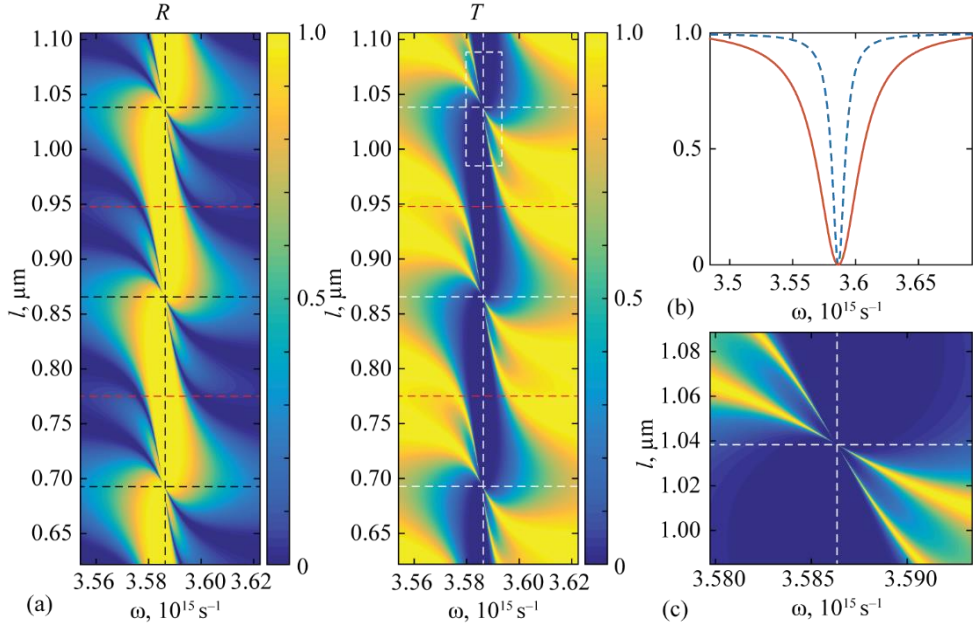

Fig. 3. (a) Reflectance $R_3(\omega, l)$ and transmittance $T_3(\omega, l)$ of the composite structure at $N = 3$ vs. the angular frequency of the incident wave and the distance $l$ between the adjacent DGs. Horizontal black and white dashed lines show the distances corresponding to the Fabry–Pérot resonances. Red dashed lines indicate the middle distances between adjacent Fabry–Pérot resonances. (b) Transmittance spectrum of the composite structure under the Fabry–Pérot resonance condition ($l = 1038$ nm) (red solid line). Dashed blue line shows the transmittance spectrum of the initial diffraction grating. (c) Magnified fragment of the transmittance in the region shown with a dashed rectangle in (a).

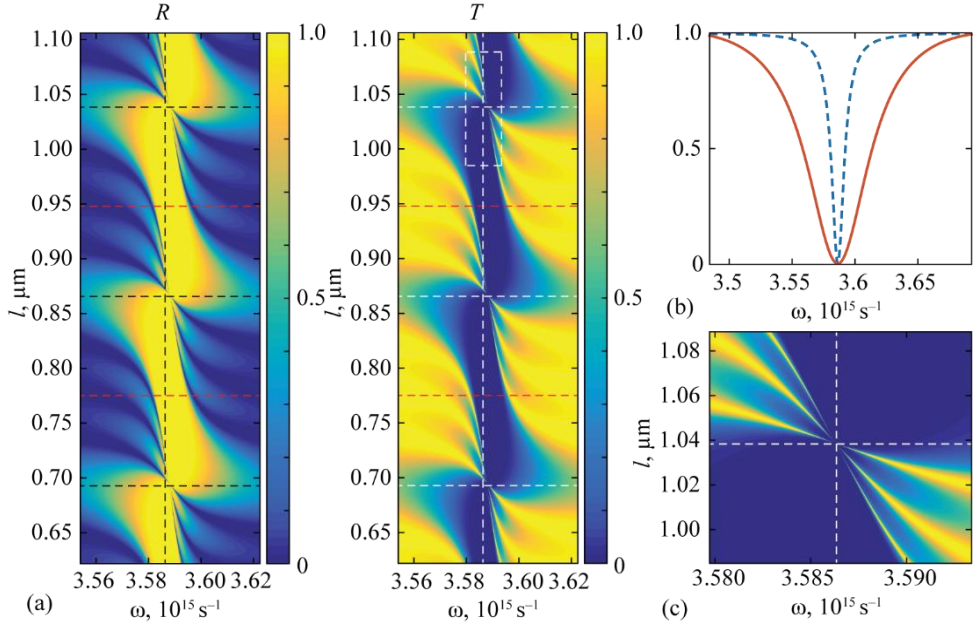

Fig. 4. (a) Reflectance $R_4(\omega, l)$ and transmittance $T_4(\omega, l)$ of the composite structure at $N = 4$ vs. the angular frequency of the incident wave and the distance $l$ between the adjacent DGs. Horizontal black and white dashed lines show the distances corresponding to the Fabry–Pérot resonances. Red dashed lines indicate the middle distances between adjacent Fabry–Pérot resonances. (b) Transmittance spectrum of the composite structure under the Fabry–Pérot resonance condition ($l = 1038$ nm) (red solid line). Dashed blue line shows the transmittance spectrum of the initial diffraction grating. (c) Magnified fragment of the transmittance in the region shown with a dashed rectangle in (a).

*5.3. Investigation and optimization of the spectra of composite structures*

In addition to the formation of a set of BICs, the spectra in Figs. 3 and 4 have another important feature, which is caused by the existence of a multiple zero of the transmission coefficient at $\omega = \omega_0 = \operatorname{Re}\omega_p$. The existence of a multiple zero makes the shape of the spectra very different from that of the initial DG, and enables obtaining bands with near-zero transmittance and near-unity reflectance centered at the frequency $\omega_0$. These bands have the most "regular" shape in the middle between two Fabry–Pérot resonances, i.e. at
$$\psi_0 + \varphi = \pi(m - 1/2),\ m \in \mathbb{N}. \tag{25}$$

The distances $l$ satisfying the condition of Eq. (25) are shown in Figs. 3 and 4 with red dashed lines. As an example, Fig. 5 shows the reflectance and transmittance spectra of the composite structures at $l = 948\,\text{nm}$ [$m = 3$ in Eq. (25)]. For comparison, dashed lines in Fig. 5 show the spectra of the initial DG. Figure 5 shows that under the condition of Eq. (25), the transmittance dip becomes closer to a rectangle with an increase in $N$. At the same time, sidelobes appear near the main transmittance dip (reflectance peak), which are especially noticeable at $N = 4$ [Fig. 5(b)]. These sidelobes are caused by the appearance of additional poles of the transmission and reflection coefficients.

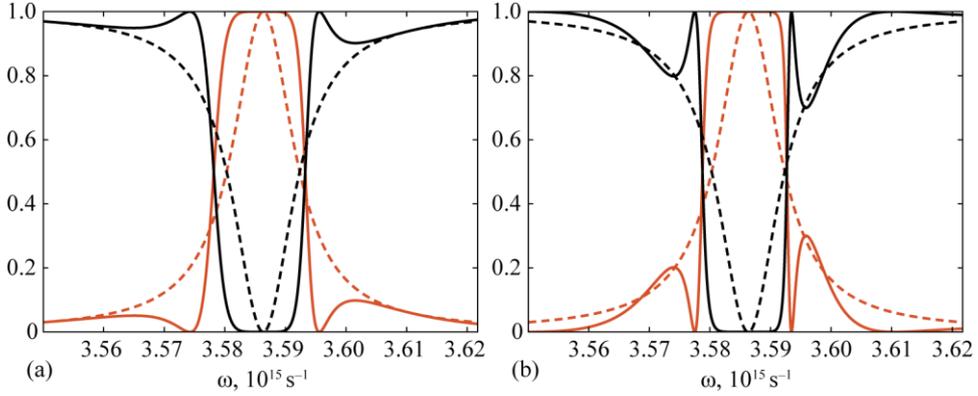

Fig. 5. Transmittance (solid black lines) and reflectance (solid red lines) of the composite structures with $N = 3$ (a) and $N = 4$ (b) at $l = 948\,\text{nm}$. Dashed lines show the transmittance (black) and reflectance (red) of the initial DG.

The existence of a multiple zero and $N$ poles in a composite structure containing $N$ diffraction gratings makes it possible to control the shape of the resonant peak (or dip) by changing the thicknesses of the dielectric layers separating the DGs. Indeed, in the general case, the poles $\omega_{p,m}$, $m = 1,...,N$ in the expression for the transmission coefficient given by Eq. (23) are the functions of the layer thicknesses $l_j$, $j = 1,...,N-1$. Above, we restricted our consideration to the case when $l_j = l$, $j = 1,...,N-1$. However, by considering the $l_j$ values as optimization parameters, one can obtain a transmittance dip with a desired shape. Let us define the required transmittance profile $T_N(\omega) = |t_N(\omega)|^2$ by the function
$$D_N(\omega) = \frac{\left[(\omega - \omega_0)/\sigma\right]^{2N}}{1 + \left[(\omega - \omega_0)/\sigma\right]^{2N}}, \tag{26}$$

which describes a smooth nearly rectangular dip with the width $2\sigma$ and no sidelobes. The representation (26) is natural for the considered composite structures, since it generalizes the Lorentzian line shape and takes into account the fact that the transmission coefficient $t_N(\omega)$ has a zero of the order $N$. It is interesting to note that the function

$$P_N(\omega) = 1 - D_N(\omega) = \frac{1}{1+\left[(\omega-\omega_0)/\sigma\right]^{2N}}, \qquad (27)$$

which describes the required reflectance peak, coincides with the squared modulus of the transfer function of the Butterworth filter of the order $N$ [22].

Let us demonstrate the possibility to control the shape of the resonant transmittance dip (reflectance peak) for the composite structures consisting of four ($N=4$) and six ($N=6$) diffraction gratings. The distance between the sidelobe maxima in the transmittance spectrum of Fig. 5(b) ($N=4$) amounts to $2\Delta\omega \approx 2.2 \cdot 10^{13}\,\text{s}^{-1}$. The spectra in Fig. 5(b) were calculated at $l_j = l = 948\,\text{nm}$, $j=1,2,3$ [$m=3$ in Eq. (25)]. Using these values as starting points, the distances $l_j$ were optimized from the condition of obtaining a transmittance spectrum described by the function $D_4(\omega)$ at $\sigma = \Delta\omega \approx 1.1 \cdot 10^{13}\,\text{s}^{-1}$. As a result of the optimization, the following values were obtained: $l_1 = l_3 = 952\,\text{nm}$ and $l_2 = 1037\,\text{nm}$. The spectra of the composite structure with $N=4$ and the optimized distances $l_j$ are shown in Fig. 6(a). Similarly, Fig. 6(b) shows the spectra of the composite structure with $N=6$ obtained by optimization with respect to the values $l_j$, $j=1,...,5$ from the condition of achieving the transmittance spectrum described by the function $D_6(\omega)$. From Fig. 6, it is evident that the obtained spectra are in good agreement with the functions $D_N(\omega)$ and $P_N(\omega)$ both at $N=4$ [Fig. 6(a)] and at $N=6$ [Fig. 6(b)].

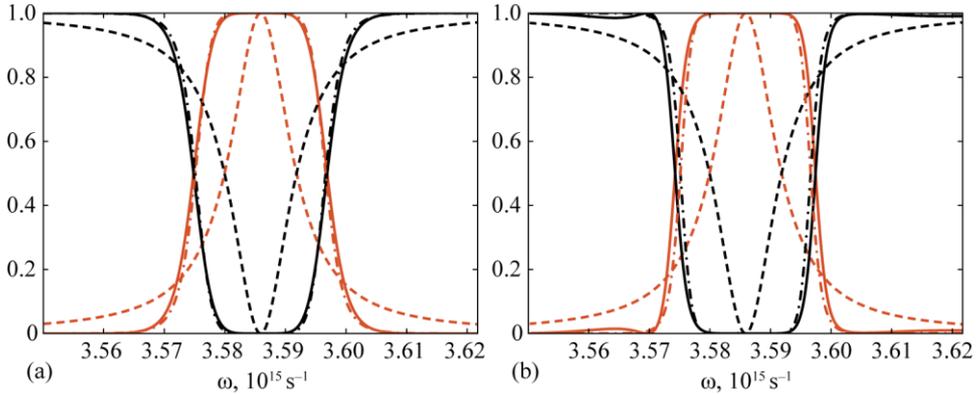

Fig. 6. Transmittance (solid black lines) and reflectance (solid red lines) of the composite structures at $N=4$ and $l_1 = l_3 = 952\,\text{nm}$, $l_2 = 1037\,\text{nm}$ (a) and at $N=6$ and $l_1 = l_5 = 1126\,\text{nm}$, $l_2 = l_4 = 1050\,\text{nm}$, $l_3 = 943\,\text{nm}$ (b). Dashed lines show the spectra of the initial DG. Dash-dot lines show the functions $D_N(\omega)$ and $P_N(\omega)$ at $N=4$ (a) and $N=6$ (b).

The width of the transmittance dips (reflectance peaks) in Fig. 6 amounts to $\Delta\omega \approx 2.3 \cdot 10^{13}\,\text{s}^{-1}$ or $\Delta\lambda \approx 3.4\,\text{nm}$. Obtaining resonant peaks or dips with an essentially subnanometer size is also of great interest. Resonant features with an arbitrarily small width can be obtained if the thicknesses of the separating layers $l_j$, $j=1,...,N$ are close to the thickness $l$ satisfying the Fabry–Pérot resonance condition of Eq. (11). As an example, Fig. 7(a) shows the transmittance of the composite structure with $N=4$ at $l_j = 1033\,\text{nm}$, $j=1,2,3$. This $l_j$ value is close to the distance $l = 1038\,\text{nm}$ satisfying the Fabry–Pérot resonance condition and shown with the upper horizontal dashed lines in Fig. 4(a). In Fig. 7(a), one can clearly see three sharp transmittance peaks against the background of a smooth minimum associated with the multiple zero of the transmission coefficient. The full widths at half maximum of the resonant peaks in Fig. 7(a) with respect to the wavelength amount to $\Delta\lambda_1 \approx 3.8\cdot 10^{-4}$ nm (left peak), $\Delta\lambda_2 \approx 3.5\cdot 10^{-3}$ nm (central peak), and $\Delta\lambda_1 \approx 5.7\cdot 10^{-2}$ nm (right peak). This effect is very similar to the electromagnetically induced transparency effect and, in the considered case, is associated with the zeros of the reflection coefficient. Let us note that the reflection coefficient of the composite structure containing four DGs ($N=4$) has four poles (which are the same as the poles of the transmission coefficient) and three zeros, which lie in the vicinity of the multiple zero of the transmission coefficient. It is interesting to mention that in the case when the condition of Eq. (11) holds, these three zeros become real, coalesce with the multiple zero of the transmission coefficient, and are cross-canceled with the real poles corresponding to the three BICs supported by the structure. This leads to the formation of a "simple" (single-pole) resonance having a Lorentzian line shape [see Fig. 4(b)].

The positions and widths of the transmittance peaks in Fig. 7(a) depend on the thicknesses $l_j$, $j=1,2,3$. Therefore, by adjusting these parameters, one can try to make the peaks coalesce, forming a single peak with a subnanometer width and a required shape. Let us demonstrate this possibility. As the desired transmittance peak shape, we chose the function $P_4(\omega)$ defined by Eq. (27) with $\sigma = 8.75\cdot 10^{10}\,\mathrm{s}^{-1}$, which is shown with dash-dot lines in Fig. 7. The peak width with respect to the wavelength amounts to $\Delta\lambda_1 \approx 2.6\cdot 10^{-2}$ nm. As a result of the optimization, the following thickness values were obtained: $l_1 = l_3 = 1027$ nm and $l_2 = 950$ nm. The transmittance spectrum of the corresponding optimized structure is shown in Fig. 7(b). The spectrum of the composite structure in Fig. 7(b) contains a single transmittance peak on an almost zero background. The shape of the obtained peak is quite close to the function $P_4(\omega)$.

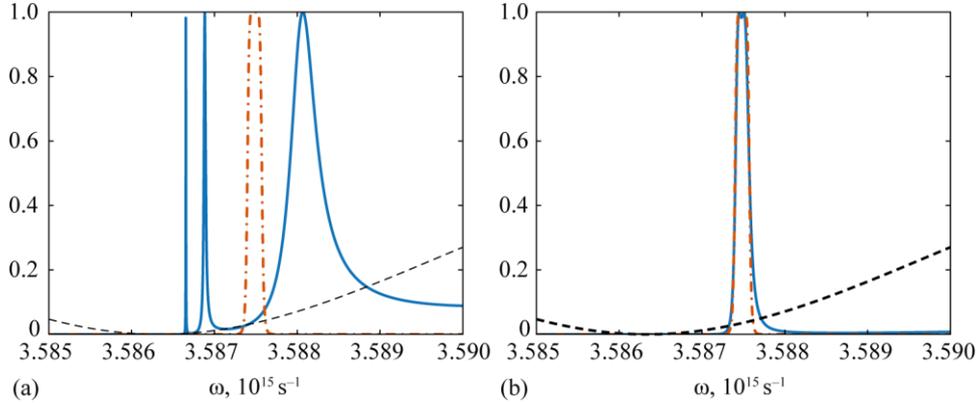

Fig. 7. Transmittance spectra (solid blue lines) of the composite structure with $N=4$ at $l_1 = l_2 = l_3 = 1033$ nm (a) and of the optimized structure with $l_1 = l_3 = 1027$ nm and $l_2 = 950$ nm. Red dash-dot lines show the desired peak shape $P_4(\omega)$. Black dashed lines show the transmittance spectrum of the initial DG.

## 6. Conclusion

In this work, using the scattering matrix formalism, we investigated resonant optical properties of composite structures consisting of several subwavelength resonant diffraction gratings separated by homogeneous layers. We demonstrated that the composite structure comprising *N* gratings has a multiple transmission zero of the order *N*. At the distance between the DGs satisfying the Fabry–Pérot resonance condition, (*N* – 1)-degenerate bound states in the continuum are formed in the composite structure. The presented theoretical results were confirmed by full-wave numerical simulations. It was shown that in the vicinity of the BICs, an effect very similar to the electromagnetically induced transparency is observed.

We also demonstrated the possibility to control the shape of the resonant transmittance dip (reflectance peak) by choosing the thicknesses of the layers separating the DGs in the composite structure. In the case when the DGs constituting the structure support a Lorentzian-shape resonance, one can obtain resonant transmittance dips (reflectance peaks) of approximately rectangular shape. The shape of the reflectance peak is well described by the transfer function of the Butterworth filter, whereas the width of the formed peak is very close to the width of the initial Lorentzian resonance supported by a single DG.

The presence of $N-1$ BICs in the composite structure provides additional degrees of freedom for controlling the shape of the spectra in the vicinity of these BICs. In particular, the presented example demonstrates that by choosing the thicknesses of the dielectric layers separating the DGs, one can obtain an approximately rectangular transmittance peak having a significantly subnanometer width.

The obtained results may find application in the design of optical filters, sensors and devices for optical differentiation and transformation of optical signals.

## Funding

## Acknowledgments

The investigation of the bound states in the continuum in the composite structures (Sections 3, 4.2) was supported by Russian Science Foundation; the studies regarding the design and investigation of the composite structures providing the required spectral peak shape (Sections 2, 4.3) were supported by Russian Foundation for Basic Research; the implementation of the simulation software and the investigation of the initial resonant grating (Section 4.1) were supported by the Russian Federation Ministry of Science and Higher Education.